\documentclass[10pt]{article}
\usepackage{multicol}
\usepackage{graphicx}
\usepackage{amsmath}
\usepackage[a4paper]{geometry}
\usepackage{rotating}

\begin{document}
\begin{center}
{\Large \bf Initial and final state temperatures of antiproton
emission sources in high energy collisions}

\vskip.75cm

Qi Wang, Fu-Hu Liu{\footnote{E-mail: fuhuliu@163.com;
fuhuliu@sxu.edu.cn}}

\vskip.25cm

{\small\it Institute of Theoretical Physics and State Key
Laboratory of Quantum Optics and Quantum Optics Devices,\\ Shanxi
University, Taiyuan, Shanxi 030006, China}

\end{center}

\vskip.5cm

{\bf Abstract:} The momentum or transverse momentum spectra of
antiprotons produced at mid-rapidity in proton-helium ($p$+He),
gold-gold (Au+Au), deuton-gold ($d$+Au), and lead-lead (Pb+Pb)
collisions over an energy range from a few GeV to a few TeV are
analyzed by the Erlang distribution, the inverse power-law (the
Hagedorn function), and the blast-wave fit, or the superposition
of two-component step function. The excitation functions of
parameters such as the mean transverse momentum, initial state
temperature, kinetic freeze-out temperature, and transverse flow
velocity increase (slightly) from a few GeV to a few TeV and from
peripheral to central collisions. At high energy and in central
collisions, large collision energy is deposited in the system,
which results in high degrees of excitation and expansion.
\\

{\bf Keywords:} Initial state temperature, final state
temperature, Erlang distribution, inverse power-law (Hagedorn
function), blast-wave fit
\\

{\bf PACS:} 12.40.Ee, 14.20.-c, 24.10.Pa, 25.75.Ag

\vskip1.0cm

\begin{multicols}{2}

{\section{Introduction}}

Temperature is an important concept in thermal and statistical
physics, high energy and nuclear physics, as well as other
scientific fields [1]. In high energy collisions, temperature is
expected to decrease from initial state to final state due to the
evolution of the interaction system. Both initial and final state
temperatures are expected to obtain from particle spectra measured
in experiments. In particular, the final state temperature is in
fact the kinetic freeze-out temperature ($T_0$) which is writhen
to transverse flow velocity ($\beta_T$) [2--4]. Both $T_0$ and
$\beta_T$ can be extracted from transverse momentum ($p_T$)
spectra of particles. It is expected that the initial state
temperature ($T_i$) can also be obtained from $p_T$ spectra [5].

Generally, to obtain temperatures such as $T_i$ and $T_0$ and
transverse flow velocity such as $\beta_T$, one should describe
$p_T$ spectra in the first place. In some cases, these parameters
are related to the models or functions which are used in the fits
to $p_T$ spectra. In other cases, these parameters are related to
the types of particles due to the non-simultaneity in the process
of particle emissions. These parameters are also related to
collision energy, system sizes, and collision centralities. The
dependences of parameters on various factors are complex. It is
very useful to find out these dependences in the understanding of
collision process.

Similar to different thermometric scales or thermometers in
thermal physics, one also expect a technical method to be used in
the description of $p_T$ spectra, which should result in some
parameters which are model or function independent. Obviously, the
average $p_T$ ($\langle p_T\rangle$) and the root-mean-square
$p_T$ ($\sqrt{\langle p_T^2\rangle}$) are model or function
independent, which are particularly determined by experimental
data themselves, though $p_T$ spectra in experiments are not
refined in full phase space. The interested parameters are
expected to relate to $\langle p_T\rangle$ and $\sqrt{\langle
p_T^2\rangle}$.

How do particles collide in high-energy collisions? What
excitation and expansion degrees of emission source can be reached
in collision process? We are interested in these and related
issues based on the particle $p_T$ spectra in experiments. It is a
novel and useful method to explore the particle collision
mechanism from the point of view of the initial state temperature
$T_i$ and the final state kinetic freeze-out temperature $T_0$ of
the emission source. Generally, transverse flow velocity $\beta_T$
is accompanied by $T_0$ in the analyses. In particular, both $T_0$
and $\beta_T$ are considerable at high energies.

There are various models or functions being used in the analyses
of $p_T$ spectra. For example, in the framework of multisource
thermal model [6--9], we could get the mean transverse momentum
$\langle p_T\rangle$ of particles and the initial state
temperature $T_i$ of emission source from fitting the $p_T$
spectra described by the Erlang distribution [7--9] which contains
the sources number $n_s$ and the mean transverse momentum $\langle
p_t\rangle$ contributed by each source. Meanwhile, we could obtain
the parameter values in the Hagedorn function [10, 11] by using
the same method, that is the method of fitting the $p_T$ spectra.
In particular, both $T_0$ and $\beta_T$ can be obtained from the
blast-wave fit [12--15] with remarkable coordination. Early
blast-wave fit is based on Boltzmann-Gibbs statistics [12--14] and
the subsequent one is based on Tsallis statistics [15].

In the rest of this paper, the Erlang distribution, the Hagedorn
function, and the blast-wave fits are given first in section 2.
Then, in section 3, these three kinds of distributions are used to
preliminarily fit the momentum and transverse momentum spectra of
antiprotons produced in high energy collisions. Several
representative groups of transverse momentum spectra are selected
to represent and summarize the changing laws of the initial and
final state temperature and other parameter values. Finally, in
section 4, we give our summary and conclusions
\\

{\section{Formalism and method}}

i) {\it The Erlang distribution}

Firstly, we discuss uniformly hard and soft collision processes in
the framework of the multisource thermal model [6--9]. According
to the model, a given particle is produced in the collision
process where a few partons have taken part in. The hard process
contains two or three partons which are valence quarks. The soft
process contains usually two or more partons which are gluons and
sea quarks. Each (the $i$-th) parton is assumed to contribute to
an exponential function [$f_i$($p_t$)] of transverse momentum
($p_t$) distribution. Let $\langle$$p_t$$\rangle$ denotes the mean
transverse momentum contributed by the $i$-th parton, we have the
probability density function of $p_t$ to be
\begin{align}
f_i(p_t)=\frac{1}{\langle p_t \rangle}
\exp\bigg(-\frac{p_t}{\langle p_t \rangle}\bigg).
\end{align}

The contribution $p_T$ of all $n_s$ partons which have taken part
in the collision process is the folding of $n_s$ exponential
functions [7--9]. We have the $p_T$ distribution $f(p_T)$ (the
probability density function of $p_T$) of final state particles to
be the Erlang distribution
\begin{align}
f_1(p_T)=\frac{1}{N}\frac{dN}{dp_T}=\frac{p_T^{n_s-1}}{(n_s-1)!{\langle
p_t \rangle}^{n}} \exp\bigg(-\frac{p_T}{\langle p_t
\rangle}\bigg),
\end{align}
which has the mean transverse momentum $\langle$$p_T$$\rangle =
n_s \langle$$p_t$$\rangle$, where $N$ denotes the number of
particles and $n_s=2$--5 in most cases.
\\

ii) {\it The Hagedorn function}

The Hagedorn function is generally suitable to describe the $p_T$
spectra of heavy flavor particles which are expectantly produced
from the hard scattering process and distributed usually in a
wider $p_T$ range. In general, the wider $p_T$ range is from 0 to
the maximum $p_T$. In refs. [10, 11], an inverse power-law results
in the probability density function of $p_T$ to be
\begin{align}
f_2(p_T)=\frac{1}{N}\frac{dN}{dp_T}=Ap_T\bigg(1+\frac{p_T}{p_0}
\bigg)^{-n},
\end{align}
where $p_0$ and $n$ are the free parameters, and $A$ is the
normalization constant. Eq. (3) is an empirical formula inspired
by quantum chromodynamics (QCD). We also call this type of inverse
power-law the Hagedorn function [10], though the inverse power-law
is more famous in the community.

In some cases, Eq. (3) is possible to describe the spectra in low
$p_T$ range which is contributed by the soft excitation process.
In fact, the spectra contributed by the hard and soft processes
represent sometimes similar trend due to the similarity which is
widely existent in high energy collisions [16--26]. Meanwhile, Eq.
(3) can be revised in different ways [27--33] which result in low
probability in low $p_T$ or high $p_T$ region by using the same or
similar parameters. We shall not discuss anymore the revisions of
Eq. (3) to avoid trivialness.
\\

iii) {\it The blast-wave fit}

We are also interested in the blast-wave fit with Boltzmann-Gibbs
statistics in its original form. According to refs. [12--14], the
blast-wave fit with Boltzmann-Gibbs statistics results in the
probability density function of $p_T$ to be
\begin{align}
f_3(p_T)=& \frac{1}{N}\frac{dN}{dp_T}=C_1p_Tm_T\int_{0}^{R}rdr\times \nonumber\\
& I_0\left[\frac{p_T\sinh
(\rho)}{T_0}\right]K_1\left[\frac{m_T\cosh (\rho)}{T_0}\right],
\end{align}
where $C_1$ is the normalized constant, $m_T = \sqrt{p_T^2 +
m_0^2}$ is the transverse mass, $m_0$ is the rest mass, $r$ is the
radial coordinate in the thermal source, $R$ is the maximum $r$
which can be regarded as the transverse size of participant, $I_0$
and $K_1$ are the modified Bessel functions of the first and
second kinds respectively, $\rho = \tanh^{-1} [\beta(r)]$ is the
boost angle, $\beta(r)=\beta_S (r/R)^{n_0}$ is a self-similar flow
profile, $\beta_S$ is the flow velocity on the surface, and
$n_0=2$ is used in the original form [12]. Generally, $\beta_T =
(2/{R^2})\int_{0}^{R}r\beta(r)dr = 2\beta_S/(n_0 + 2)$.

According to ref. [15], the blast-wave fit with Tsallis statistics
results in the probability density function of $p_T$ to be
\begin{align}
f_4(p_T)=&
\frac{1}{N}\frac{dN}{dp_T}=C_2p_Tm_T\int_{-\pi}^{\pi}d\phi
\int_{0}^{R}rdr \Big \{1+  \nonumber\\
&  \frac{q-1}{T_0} \big [m_T\cosh (\rho)-p_T \sinh (\rho) \cos
(\phi) \big ] \Big \}^{{-1}/{(q-1)}},
\end{align}
where $C_2$ is the normalized constant, $q$ is an entropy index
that characterizes the degree of non-equilibrium, $\phi$ denotes
the azimuthal angle, and $n_0=1$ is used in the original form
[15]. Because of $n_0$ being an insensitive quantity, the results
corresponding to $n_0=1$ and 2 for the blast-wave model with
Boltzmann-Gibbs or Tsallis statistics are harmonious [34]. In
addition, the index $-1/(q-1)$ used in Eq. (5) can be replaced by
$-q/(q - 1)$ due to $q$ being very close to 1. This substitution
results in a small and negligible difference in the distribution

As we know, the blast-wave fit with Boltzmann-Gibbs statistics is
consistent to the fit with Tsallis statistics. In many cases, one
of them is enough to use in the dada analysis. In this work, we
use only the blast-wave fit with Tsallis statistics, though the
fit with Boltzmann-Gibbs statistics is also usable and acceptable.
\\

iv) {\it Monte Carlo calculation based on $p_T$ distribution}

Based on one of probability density functions of $p_T$ discussed
above and the assumption of isotropic emission in sources rest
frame, we can obtain other quantities and distributions. In
particular, if the analytic expression is difficult to obtain, we
can use the Monte Carlo method to obtain some concerned
quantities, and the distributions of these concerned quantities
can be obtained by statistics. These concerned quantities include,
but are not limited to, momentum, energy, rapidity, velocity, and
others. Conversely, the concerned $p_T$ and its distribution can
be obtained from other distribution and the assumption of
isotropic emission in sources rest frame.

In the Monte Carlo method [35], let $R_{1,2}$ denote random
numbers distributed evenly in [0,1]. Some discrete values of $p_T$
can be obtained due to the following limitation
\begin{align}
\int_0^{p_T}f_{p_T}(p'_T)dp'_T <R_1 <\int_0^{p_T+\delta
p_T}f_{p_T}(p'_T) dp'_T,
\end{align}
where $\delta p_T$ denote a small shift relative to $p_T$. When
the reference frame is transformed along the longitudinal
direction ($z$-axis), $p_T$ and its distribution do not changed
determinately.

The change of rapidity in the transformation of reference frame
should be satisfied
\begin{align}
y=y'+R_2 (y_{\max} -y_{\min})+y_{\min},
\end{align}
where $y'$ denotes the rapidity of concerned particle in the
sources rest frame, which can be obtained from the discrete values
of $p_T$ and the assumption of isotropic emission, $y$ is the
rapidity after the transformation of reference frame, and
$y_{\max}$ and $y_{\min}$ are the maximum and minimum rapidity
shifts of the source before the transformation of reference frame.
After the transformation of reference frame, the $z$-component
($p_z$) of momentum ($p$) can be given by
\begin{align}
p_z =\sqrt{p_T^2+m_0^2}\sinh(y).
\end{align}
Naturally,
\begin{align}
p=\sqrt{p_z^2+p_T^2}.
\end{align}

According to refs. [36--38], the initial temperature is determined
by
\begin{align}
T_i =\sqrt{\frac{\langle p_T^2 \rangle}{2}}.
\end{align}
If the $x$-component ($p_x$) and $y$-component ($p_y$) of $p$ are
considered before or after the transformation of reference frame,
we have
\begin{align}
T_i =\sqrt{\langle p_x^2 \rangle}=\sqrt{\langle p_y^2 \rangle}.
\end{align}
Obviously, $T_i$ is invariant in the transformation of reference
frame. In the sources rest frame, if the $z$-component of momentum
is $p_z'$, we also have
\begin{align}
T_i =\sqrt{\langle p_z'^2 \rangle}
\end{align}
due to the assumption of isotropic emission.
\\

\begin{figure*}[!htb]
\begin{center}
\includegraphics[width=16cm]{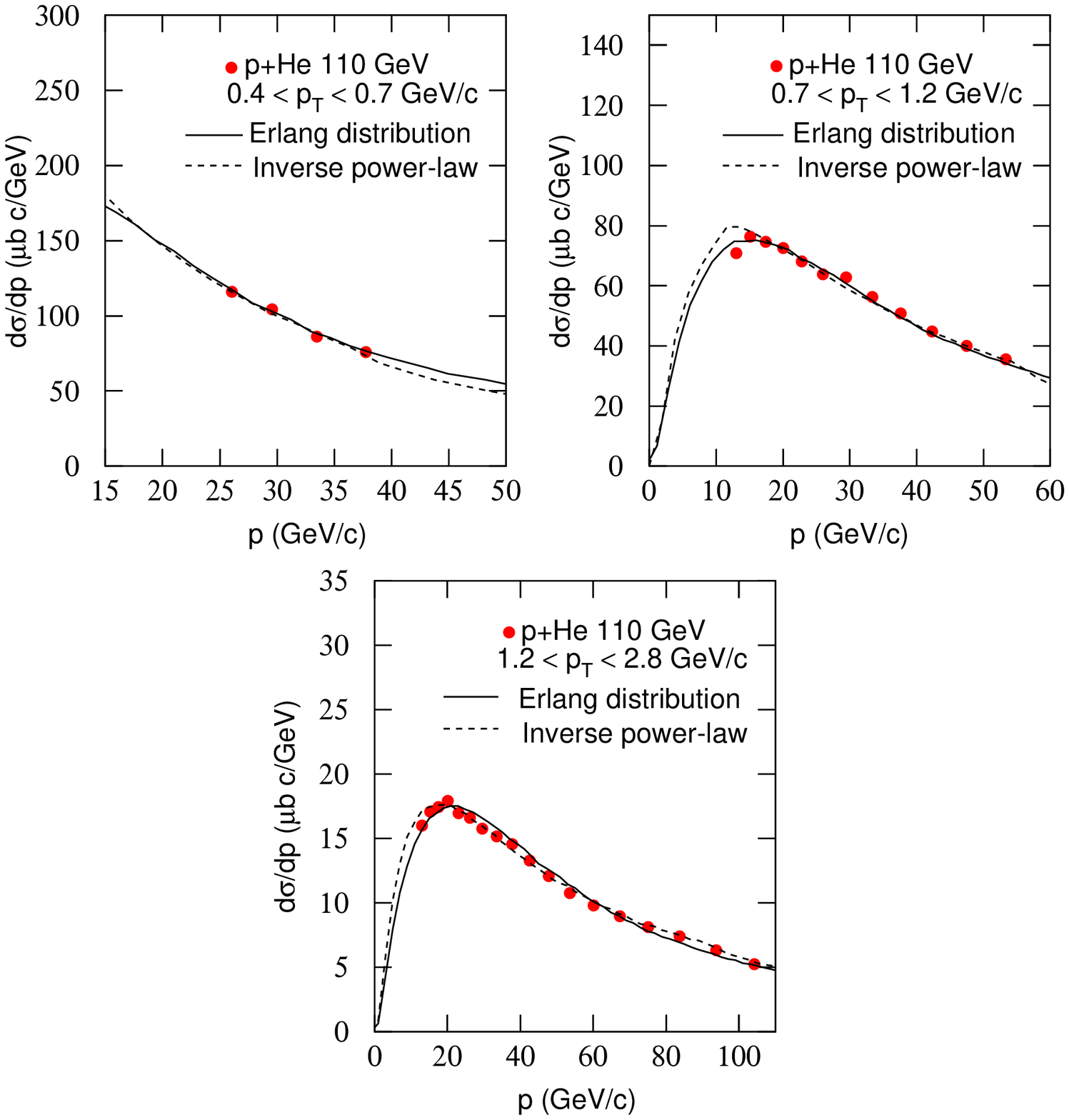}
\end{center}
{\small Fig. 1. Antiproton production differential cross-section
as a function of momentum, integrated over various transverse
momentum ranges, in $p$+He collisions at 110 GeV. The data
represent the results of LHCb Collaboration [39]. The solid and
dashed curves represent the results fitted by the Erlang
distribution Eq. (2) and the inverse power-law Eq. (3), where the
Monte Carlo calculation is used to transform transverse momenta to
momenta.}
\end{figure*}

\begin{figure*}[!htb]
\begin{center}
\includegraphics[width=16cm]{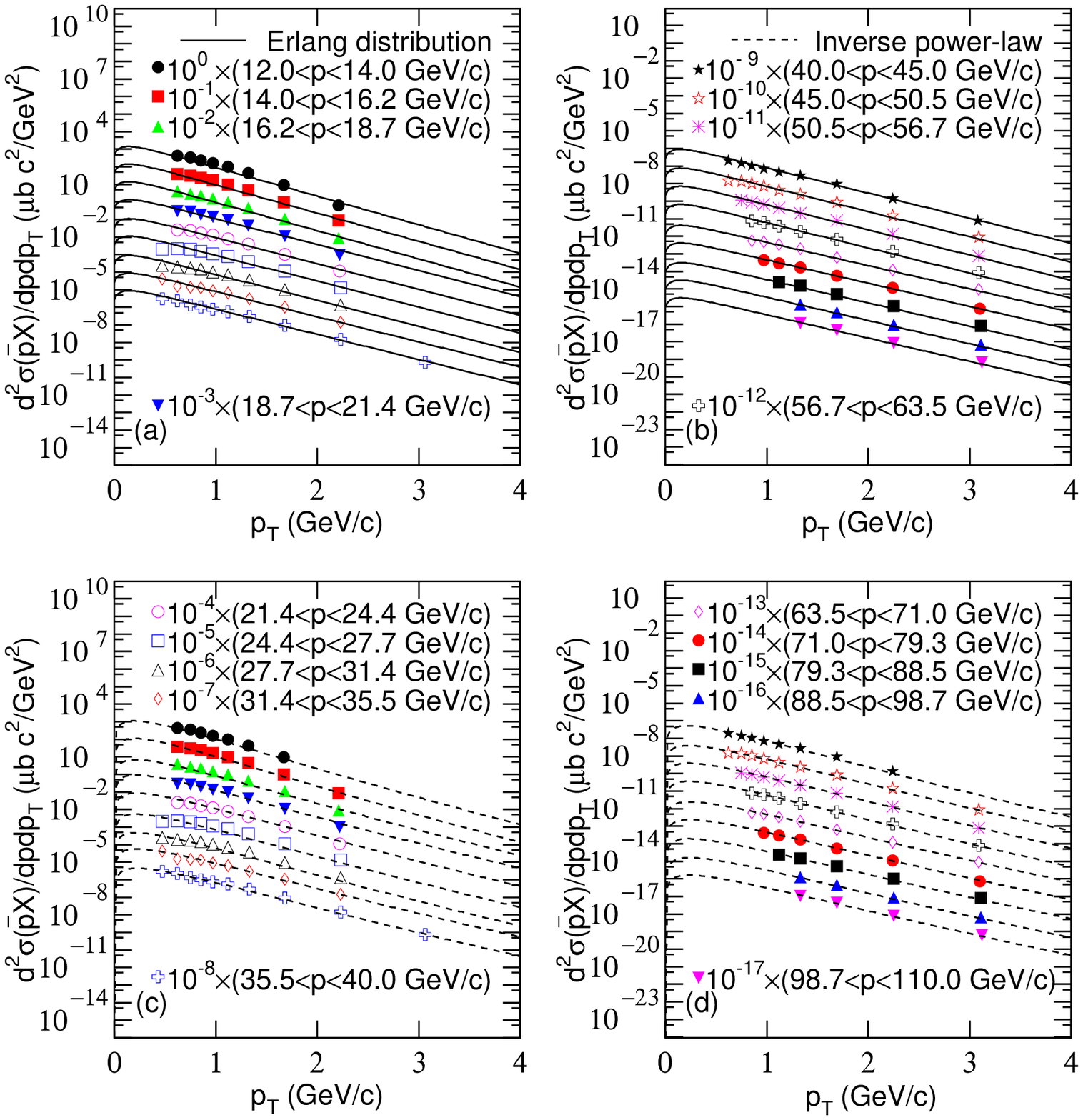}
\end{center}
{\small Fig. 2. Antiproton production double differential
cross-section as a function of transverse momentum, integrated
over various momentum ranges and divided by the ranges, in $p$+He
collisions at 110 GeV. The data represent the results of LHCb
Collaboration [40]. The solid and dashed curves represent the
results fitted by the Erlang distribution Eq. (2) and the inverse
power-law Eq. (3), where the Monte Carlo calculation is used to
transform transverse momenta to momenta.}
\end{figure*}

\begin{figure*}[!htb]
\begin{center}
\includegraphics[width=16cm]{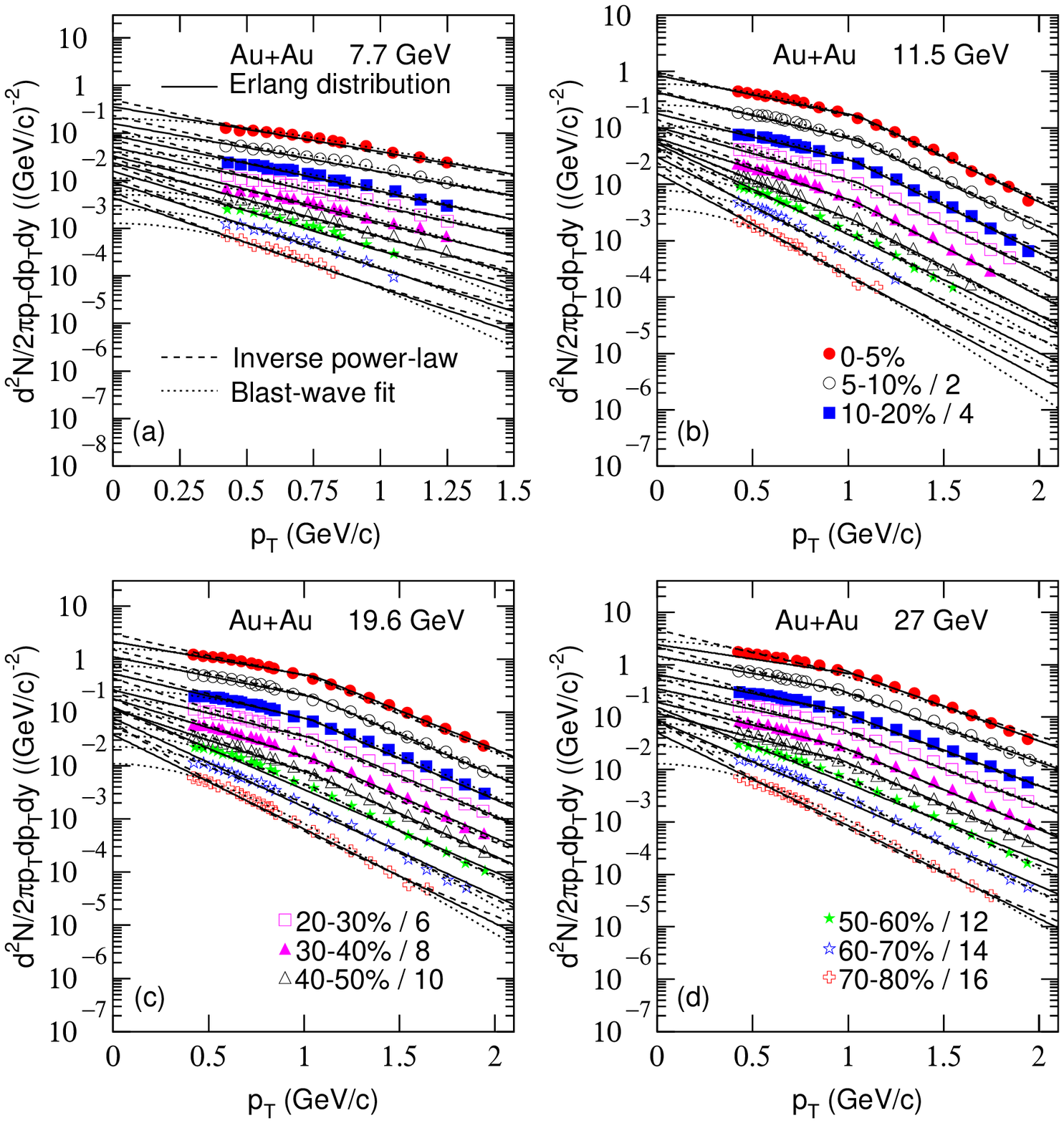}
\end{center}
{\small Fig. 3. Transverse momentum spectra, $d^2N/2\pi
p_Tdp_Tdy$, of $\bar p$ produced in mid-rapidity $|y|<0.5$ in
Au+Au collisions at four energies: (a) 7.7, (b) 11.5, (c) 19.6 and
(d) 27 GeV. The symbols represent the experimental data measured
by the STAR Collaboration in different centrality intervals [41],
which are scaled by different amounts marked in the panels. The
solid, dashed, and dotted curves are our results fitted by using
the Erlang distribution Eq. (2), the inverse power-law Eq. (3),
and the blast-wave fit Eq. (5), respectively.}
\end{figure*}

\end{multicols}
\begin{sidewaystable}
{\scriptsize Table 1. Values of $\langle p_T\rangle$, $p_0$, $n$,
and $\chi^2$/dof corresponding to the solid and dashed curves in
Figs. 1 and 2, where the values of $\chi^2$/dof for the solid and
dashed curves are listed in terms of value$_1$/value$_2$.}
\vspace{-4mm}
\begin{center}
{\tiny
\begin{tabular} {ccccccccccc}\\ \hline\hline Figure & Range of $p$ ($p_T$) (GeV/$c$) & $\langle p_T\rangle$ (GeV/$c$) & $p_0$ (GeV/$c$) & $n$ & $\chi^2$/dof\\
\hline
Figure 1       & 0.4--0.7 & $3.000\pm0.020$ & $56.30\pm0.018$ & $12.30\pm0.30$ & $0.597/0.981$\\
               & 0.7--1.2 & $4.500\pm0.013$ & $14.50\pm0.010$ & $4.20\pm0.10$  & $1.166/3.017$\\
               & 1.2--2.8 & $6.460\pm0.020$ & $9.80\pm0.024$  & $5.00\pm0.10$  & $0.994/1.489$\\
\hline
Figure 2       & 12.0--14.0 & $0.961\pm0.005$ & $9.00\pm0.02$ & $48.00\pm0.01$ & $0.257/0.628$\\
               & 14.0--16.2 & $0.973\pm0.006$ & $9.00\pm0.01$ & $48.00\pm0.01$ & $0.125/0.252$\\
               & 16.2--18.7 & $0.986\pm0.006$ & $9.00\pm0.02$ & $48.00\pm0.01$ & $0.100/0.845$\\
               & 18.7--21.4 & $1.003\pm0.008$ & $9.00\pm0.02$ & $48.00\pm0.01$ & $0.210/0.572$\\
               & 21.4--24.4 & $1.011\pm0.004$ & $9.00\pm0.01$ & $48.00\pm0.01$ & $0.213/0.270$\\
               & 24.4--27.7 & $0.943\pm0.005$ & $9.00\pm0.01$ & $48.00\pm0.01$ & $0.090/0.734$\\
               & 27.7--31.4 & $0.943\pm0.010$ & $9.00\pm0.02$ & $48.00\pm0.01$ & $0.134/0.733$\\
               & 31.4--35.5 & $0.943\pm0.008$ & $9.00\pm0.02$ & $48.00\pm0.01$ & $0.376/0.236$\\
               & 35.5--40.0 & $0.987\pm0.007$ & $9.00\pm0.01$ & $48.00\pm0.01$ & $0.196/0.098$\\
               & 40.0--45.0 & $1.109\pm0.005$ & $9.00\pm0.01$ & $48.00\pm0.02$ & $0.221/0.073$\\
               & 45.0--50.5 & $1.109\pm0.010$ & $9.00\pm0.02$ & $48.00\pm0.02$ & $0.133/0.084$\\
               & 50.5--56.7 & $1.199\pm0.010$ & $9.00\pm0.01$ & $48.00\pm0.01$ & $0.063/0.037$\\
               & 56.7--63.5 & $1.288\pm0.015$ & $9.00\pm0.02$ & $48.00\pm0.01$ & $0.136/0.053$\\
               & 63.5--71.0 & $1.301\pm0.015$ & $9.00\pm0.01$ & $48.00\pm0.01$ & $0.162/0.106$\\
               & 71.0--79.3 & $1.407\pm0.020$ & $9.00\pm0.01$ & $48.00\pm0.01$ & $0.113/0.125$\\
               & 79.3--88.5 & $1.541\pm0.020$ & $9.00\pm0.02$ & $48.00\pm0.01$ & $0.113/0.153$\\
               & 88.5--98.7 & $1.729\pm0.020$ & $9.00\pm0.01$ & $48.00\pm0.01$ & $0.190/0.065$\\
               & 98.7--110.0 & $1.739\pm0.015$ & $9.00\pm0.01$ & $48.00\pm0.01$& $0.225/0.425$\\
\hline
\end{tabular}}
\end{center}
\end{sidewaystable}

\begin{sidewaystable}
{\scriptsize Table 2. Values of $\langle p_T\rangle$, $T_i$,
$T_0$, $\beta_T$, $p_0$, and $n$ corresponding to the curves in
Figs. 3--5. The values presented in terms of value$_1$/value$_2$
denote respectively the parameters of the first and second
components. In all cases, $q=1.01$ which is not listed in the
table. The values of $\chi^2$/dof are in normal range and not
listed in the table to reduce its length.} \vspace{-4mm}
\begin{center}
{\tiny
\begin{tabular} {ccccccccccc}\\ \hline\hline Collisions & Centrality & $\langle p_T\rangle$ (GeV/$c$) & $T_i$ (GeV) & $T_0$ (GeV) & $\beta_T$ ($c$) & $p_0$ (GeV/$c$) & $n$ \\
\hline
7.7 GeV Au+Au & 0--5\%   & $0.785\pm0.023$ & $0.579\pm0.018$ & $0.135\pm0.002$ & $0.340\pm0.020$ & $3.80\pm0.20$ & $11.10\pm0.40$\\
              & 5--10\%  & $0.777\pm0.013$ & $0.573\pm0.010$ & $0.133\pm0.003$ & $0.325\pm0.025$ & $3.60\pm0.30$ & $11.25\pm0.75$\\
              & 10--20\% & $0.762\pm0.030$ & $0.562\pm0.024$ & $0.126\pm0.003$ & $0.295\pm0.015$ & $3.55\pm0.20$ & $12.50\pm0.50$\\
              & 20--30\% & $0.754\pm0.025$ & $0.556\pm0.020$ & $0.125\pm0.003$ & $0.285\pm0.015$ & $3.47\pm0.35$ & $13.00\pm1.00$\\
              & 30--40\% & $0.732\pm0.020$ & $0.541\pm0.016$ & $0.122\pm0.004$ & $0.255\pm0.015$ & $3.35\pm0.35$ & $13.50\pm0.50$\\
              & 40--50\% & $0.718\pm0.025$ & $0.530\pm0.014$ & $0.116\pm0.003$ & $0.235\pm0.015$ & $3.25\pm0.20$ & $14.30\pm0.70$\\
              & 50--60\% & $0.664\pm0.020$ & $0.484\pm0.015$ & $0.108\pm0.003$ & $0.175\pm0.025$ & $3.20\pm0.10$ & $15.00\pm0.30$\\
              & 60--70\% & $0.653\pm0.015$ & $0.477\pm0.014$ & $0.106\pm0.004$ & $0.165\pm0.015$ & $3.00\pm0.10$ & $15.40\pm0.30$\\
              & 70--80\% & $0.590\pm0.010$ & $0.425\pm0.018$ & $0.096\pm0.004$ & $0.145\pm0.015$ & $2.90\pm0.20$ & $15.50\pm0.50$\\
\hline
11.5 GeV Au+Au & 0--5\%   & $0.728\pm0.020/1.318\pm0.105$ & $0.530\pm0.015/0.945\pm0.035$ & $0.185\pm0.005/0.113\pm0.013$ & $0.315\pm0.005/0.280\pm0.010$ & $4.30\pm0.20/3.10\pm0.10$ & $8.50\pm0.50/16.00\pm0.50$\\
               & 5--10\%  & $0.722\pm0.031/1.319\pm0.098$ & $0.526\pm0.018/0.945\pm0.025$ & $0.180\pm0.015/0.108\pm0.012$ & $0.314\pm0.010/0.275\pm0.020$ & $4.25\pm0.20/3.10\pm0.10$ & $9.00\pm0.50/16.50\pm0.50$\\
               & 10--20\% & $0.723\pm0.028/1.312\pm0.151$ & $0.527\pm0.030/0.940\pm0.050$ & $0.170\pm0.010/0.105\pm0.005$ & $0.314\pm0.010/0.268\pm0.013$ & $4.20\pm0.30/3.10\pm0.10$ & $10.00\pm0.50/17.00\pm0.50$\\
               & 20--30\% & $0.716\pm0.018/1.298\pm0.168$ & $0.515\pm0.025/0.929\pm0.085$ & $0.127\pm0.010              $ & $0.255\pm0.010              $ & $4.00\pm0.30/3.10\pm0.20$ & $11.00\pm0.50/17.00\pm0.50$\\
               & 30--40\% & $0.701\pm0.020/1.277\pm0.155$ & $0.510\pm0.040/0.911\pm0.082$ & $0.124\pm0.010              $ & $0.240\pm0.010              $ & $3.70\pm0.30/3.10\pm0.30$ & $12.15\pm0.50/17.50\pm0.50$\\
               & 40--50\% & $0.684\pm0.035/1.257\pm0.121$ & $0.496\pm0.046/0.893\pm0.082$ & $0.119\pm0.008              $ & $0.228\pm0.012              $ & $3.45\pm0.15/3.10\pm0.10$ & $12.65\pm0.60/18.10\pm0.20$\\
               & 50--60\% & $0.697\pm0.010              $ & $0.531\pm0.018              $ & $0.113\pm0.008              $ & $0.218\pm0.012              $ & $3.30\pm0.20/3.20\pm0.10$ & $12.75\pm0.75/18.20\pm0.20$\\
               & 60--70\% & $0.679\pm0.010              $ & $0.504\pm0.018              $ & $0.110\pm0.008              $ & $0.205\pm0.015              $ & $3.20\pm0.10            $ & $15.90\pm0.30             $\\
               & 70--80\% & $0.656\pm0.010              $ & $0.483\pm0.018              $ & $0.105\pm0.008              $ & $0.185\pm0.015              $ & $3.00\pm0.10            $ & $16.25\pm0.75             $\\
\hline
19.6 Gev Au+Au & 0--5\%   & $0.735\pm0.021/1.336\pm0.125$ & $0.535\pm0.015/0.958\pm0.037$ & $0.190\pm0.010/0.110\pm0.010$ & $0.330\pm0.010/0.300\pm0.020$ & $4.50\pm0.20/3.40\pm0.20$ & $9.00\pm0.50/16.50\pm0.50$\\
               & 5--10\%  & $0.727\pm0.020/1.326\pm0.124$ & $0.530\pm0.010/0.949\pm0.030$ & $0.185\pm0.010/0.107\pm0.008$ & $0.315\pm0.005/0.300\pm0.010$ & $4.40\pm0.20/3.30\pm0.20$ & $9.50\pm1.00/17.40\pm0.60$\\
               & 10--20\% & $0.721\pm0.010/1.325\pm0.125$ & $0.525\pm0.015/0.950\pm0.040$ & $0.184\pm0.010/0.105\pm0.005$ & $0.313\pm0.010/0.300\pm0.013$ & $4.20\pm0.20/3.40\pm0.20$ & $10.50\pm0.50/17.40\pm0.60$\\
               & 20--30\% & $0.719\pm0.019/1.324\pm0.176$ & $0.523\pm0.013/0.950\pm0.080$ & $0.132\pm0.004              $ & $0.280\pm0.010              $ & $4.00\pm0.40/3.30\pm0.20$ & $11.10\pm0.50/17.40\pm0.90$\\
               & 30--40\% & $0.700\pm0.100/1.323\pm0.211$ & $0.510\pm0.045/0.949\pm0.079$ & $0.128\pm0.003              $ & $0.255\pm0.005              $ & $3.80\pm0.20/3.20\pm0.30$ & $12.00\pm0.80/17.80\pm0.60$\\
               & 40--50\% & $0.694\pm0.090/1.319\pm0.209$ & $0.506\pm0.072/0.945\pm0.120$ & $0.123\pm0.003              $ & $0.253\pm0.005              $ & $3.60\pm0.20/3.10\pm0.20$ & $12.60\pm0.50/17.80\pm0.20$\\
               & 50--60\% & $0.737\pm0.010              $ & $0.569\pm0.018              $ & $0.118\pm0.003              $ & $0.235\pm0.005              $ & $3.40\pm0.20/3.00\pm0.10$ & $12.60\pm0.40/18.20\pm0.50$\\
               & 60--70\% & $0.705\pm0.015              $ & $0.542\pm0.014              $ & $0.113\pm0.003              $ & $0.205\pm0.005              $ & $3.40\pm0.20/3.00\pm0.20$ & $12.90\pm0.90/19.00\pm0.30$\\
               & 70--80\% & $0.676\pm0.015              $ & $0.516\pm0.014              $ & $0.110\pm0.002              $ & $0.185\pm0.005              $ & $3.10\pm0.10            $ & $16.90\pm0.50             $\\
\hline
27 GeV Au+Au & 0--5\%   & $0.726\pm0.028/1.328\pm0.153$ & $0.528\pm0.015/0.954\pm0.052$ & $0.145\pm0.003$ & $0.300\pm0.005$ & $4.90\pm0.20/3.90\pm0.20$ & $10.50\pm1.20/16.50\pm0.50$\\
             & 5--10\%  & $0.720\pm0.022/1.329\pm0.108$ & $0.523\pm0.019/0.955\pm0.020$ & $0.142\pm0.003$ & $0.300\pm0.005$ & $4.70\pm0.20/3.60\pm0.20$ & $10.50\pm1.00/16.50\pm0.50$\\
             & 10--20\% & $0.719\pm0.029/1.323\pm0.143$ & $0.522\pm0.011/0.950\pm0.023$ & $0.138\pm0.002$ & $0.295\pm0.005$ & $4.50\pm0.20/3.70\pm0.20$ & $11.25\pm0.50/16.50\pm0.50$\\
             & 20--30\% & $0.714\pm0.025/1.318\pm0.188$ & $0.518\pm0.010/0.946\pm0.035$ & $0.136\pm0.003$ & $0.285\pm0.005$ & $4.30\pm0.30/3.60\pm0.20$ & $11.25\pm0.75/17.10\pm0.50$\\
             & 30--40\% & $0.708\pm0.035/1.303\pm0.203$ & $0.515\pm0.025/0.936\pm0.058$ & $0.131\pm0.003$ & $0.269\pm0.002$ & $4.10\pm0.20/3.50\pm0.20$ & $11.75\pm0.75/17.30\pm0.50$\\
             & 40--50\% & $0.706\pm0.065/1.296\pm0.250$ & $0.513\pm0.040/0.930\pm0.085$ & $0.128\pm0.003$ & $0.263\pm0.003$ & $3.90\pm0.20/3.30\pm0.20$ & $13.00\pm0.50/17.70\pm0.30$\\
             & 50--60\% & $0.760\pm0.020              $ & $0.580\pm0.018              $ & $0.125\pm0.003$ & $0.225\pm0.003$ & $3.90\pm0.20/3.20\pm0.10$ & $13.00\pm0.30/18.20\pm0.40$\\
             & 60--70\% & $0.730\pm0.025              $ & $0.564\pm0.014              $ & $0.123\pm0.003$ & $0.205\pm0.005$ & $3.70\pm0.30/3.00\pm0.20$ & $14.00\pm0.50/18.90\pm0.70$\\
             & 70--80\% & $0.722\pm0.022              $ & $0.521\pm0.014              $ & $0.118\pm0.002$ & $0.179\pm0.003$ & $3.20\pm0.10            $ & $17.50\pm0.50             $\\
\hline
39 GeV Au+Au & 0--5\%   & $0.740\pm0.021/1.352\pm0.132$ & $0.538\pm0.014/0.970\pm0.030$ & $0.195\pm0.015/0.115\pm0.005$ & $0.360\pm0.010/0.315\pm0.017$ & $5.30\pm0.20/4.20\pm0.20$ & $8.75\pm0.75/17.00\pm1.00 $\\
             & 5--10\%  & $0.738\pm0.019/1.343\pm0.128$ & $0.537\pm0.010/0.963\pm0.030$ & $0.194\pm0.020/0.115\pm0.003$ & $0.355\pm0.010/0.315\pm0.018$ & $5.10\pm0.20/4.00\pm0.20$ & $9.00\pm0.50/17.00\pm0.50 $\\
             & 10--20\% & $0.736\pm0.018/1.338\pm0.139$ & $0.535\pm0.010/0.959\pm0.022$ & $0.193\pm0.010/0.115\pm0.004$ & $0.350\pm0.010/0.315\pm0.015$ & $4.90\pm0.20/3.80\pm0.20$ & $10.00\pm0.50/17.00\pm1.00$\\
             & 20--30\% & $0.727\pm0.019/1.329\pm0.156$ & $0.529\pm0.009/0.953\pm0.028$ & $0.137\pm0.002              $ & $0.294\pm0.004              $ & $4.70\pm0.20/3.70\pm0.20$ & $11.00\pm0.50/17.50\pm1.00$\\
             & 30--40\% & $0.718\pm0.010/1.329\pm0.118$ & $0.523\pm0.013/0.953\pm0.040$ & $0.135\pm0.002              $ & $0.283\pm0.003              $ & $4.50\pm0.30/3.60\pm0.20$ & $11.25\pm0.75/17.40\pm0.60$\\
             & 40--50\% & $0.837\pm0.040              $ & $0.600\pm0.022              $ & $0.133\pm0.003              $ & $0.255\pm0.005              $ & $4.30\pm0.20/3.35\pm0.25$ & $13.50\pm0.50/17.50\pm1.00$\\
             & 50--60\% & $0.813\pm0.020              $ & $0.580\pm0.025              $ & $0.130\pm0.003              $ & $0.233\pm0.003              $ & $4.00\pm0.20/3.20\pm0.10$ & $14.20\pm0.40/18.30\pm0.30$\\
             & 60--70\% & $0.776\pm0.025              $ & $0.569\pm0.022              $ & $0.128\pm0.002              $ & $0.213\pm0.003              $ & $3.80\pm0.20/3.10\pm0.10$ & $14.60\pm0.20/19.00\pm0.50$\\
             & 70--80\% & $0.750\pm0.022              $ & $0.547\pm0.028              $ & $0.125\pm0.002              $ & $0.193\pm0.003              $ & $3.38\pm0.13            $ & $17.50\pm0.20             $\\
\hline
\end{tabular}}
\end{center}
\end{sidewaystable}

\begin{sidewaystable}
{\scriptsize Table 2. Continued.} \vspace{-4mm}
\begin{center}
{\tiny
\begin{tabular} {ccccccccccc}\\ \hline\hline Collisions & Centrality & $\langle p_T\rangle$ (GeV/$c$) & $T_i$ (GeV) & $T_0$ (GeV) & $\beta_T$ ($c$) & $p_0$ (GeV/$c$) & $n$ \\
\hline
62.4 GeV Au+Au & 0--5\%   & $1.130\pm0.025$ & $0.856\pm0.020$ & $0.210\pm0.010$ & $0.425\pm0.010$ & $8.10\pm0.30/11.00\pm0.50$ & $6.00\pm0.50/21.00\pm1.00 $\\
               & 5--10\%  & $1.110\pm0.015$ & $0.842\pm0.012$ & $0.205\pm0.005$ & $0.405\pm0.010$ & $7.80\pm0.30/10.50\pm0.50$ & $7.00\pm0.50/21.00\pm1.00 $\\
               & 10--20\% & $1.104\pm0.015$ & $0.839\pm0.012$ & $0.195\pm0.005$ & $0.395\pm0.005$ & $7.50\pm0.30/10.50\pm0.30$ & $8.00\pm0.30/21.00\pm1.00 $\\
               & 20--30\% & $1.063\pm0.035$ & $0.811\pm0.029$ & $0.178\pm0.003$ & $0.370\pm0.010$ & $7.20\pm0.30/10.00\pm0.50$ & $9.00\pm0.50/22.00\pm2.00 $\\
               & 30--40\% & $1.022\pm0.035$ & $0.782\pm0.029$ & $0.162\pm0.004$ & $0.350\pm0.010$ & $6.80\pm0.40/9.00\pm0.50$  & $10.00\pm0.50/22.50\pm1.50$\\
               & 40--50\% & $0.989\pm0.015$ & $0.758\pm0.012$ & $0.155\pm0.005$ & $0.310\pm0.010$ & $6.50\pm0.30/7.60\pm0.40$  & $11.00\pm0.50/22.00\pm1.00$\\
               & 50--60\% & $0.933\pm0.010$ & $0.718\pm0.008$ & $0.150\pm0.002$ & $0.295\pm0.005$ & $6.65\pm0.25$              & $16.50\pm0.50$\\
               & 60--70\% & $0.871\pm0.015$ & $0.671\pm0.012$ & $0.147\pm0.002$ & $0.265\pm0.005$ & $6.30\pm0.30$              & $17.75\pm0.75$\\
               & 70--80\% & $0.840\pm0.012$ & $0.648\pm0.013$ & $0.143\pm0.003$ & $0.215\pm0.005$ & $6.00\pm0.30$              & $18.50\pm1.00$\\
\hline
130 GeV Au+Au & 0--6\%   & $1.139\pm0.025$ & $0.862\pm0.020$ & $0.220\pm0.010$ & $0.435\pm0.010$ & $11.70\pm0.30$ & $13.50\pm1.50$\\
              & 6--11\%  & $1.125\pm0.025$ & $0.853\pm0.020$ & $0.210\pm0.005$ & $0.410\pm0.010$ & $11.20\pm0.30$ & $13.70\pm1.50$\\
              & 12--18\% & $1.115\pm0.015$ & $0.846\pm0.012$ & $0.198\pm0.005$ & $0.398\pm0.008$ & $10.70\pm0.30$ & $13.90\pm1.10$\\
              & 18--26\% & $1.104\pm0.015$ & $0.839\pm0.012$ & $0.187\pm0.007$ & $0.380\pm0.005$ & $10.35\pm0.50$ & $14.20\pm2.00$\\
              & 26--34\% & $1.089\pm0.020$ & $0.828\pm0.012$ & $0.165\pm0.005$ & $0.365\pm0.010$ & $10.15\pm0.20$ & $15.50\pm1.30$\\
              & 34--45\% & $1.051\pm0.015$ & $0.802\pm0.012$ & $0.160\pm0.005$ & $0.360\pm0.010$ & $10.00\pm0.40$ & $16.50\pm1.50$\\
              & 45--58\% & $1.001\pm0.015$ & $0.767\pm0.012$ & $0.150\pm0.005$ & $0.350\pm0.010$ & $9.00\pm0.20 $ & $17.80\pm1.30$\\
              & 58--85\% & $0.941\pm0.020$ & $0.724\pm0.010$ & $0.143\pm0.003$ & $0.315\pm0.005$ & $8.50\pm0.20 $ & $19.50\pm1.20$\\
\hline
200 GeV Au+Au & 0--5\%   & $1.141\pm0.010$ & $0.864\pm0.008$ & $0.220\pm0.020$ & $0.435\pm0.010$ & $12.60\pm0.30/12.00\pm0.50$ & $10.00\pm0.50/18.00\pm2.00$\\
              & 5--10\%  & $1.125\pm0.015$ & $0.854\pm0.012$ & $0.210\pm0.010$ & $0.415\pm0.010$ & $12.30\pm0.30/11.00\pm0.50$ & $11.00\pm2.00/18.00\pm5.00$\\
              & 10--20\% & $1.118\pm0.025$ & $0.848\pm0.020$ & $0.200\pm0.015$ & $0.410\pm0.015$ & $12.00\pm0.30/10.50\pm0.50$ & $12.50\pm0.50/18.00\pm2.00$\\
              & 20--30\% & $1.095\pm0.025$ & $0.832\pm0.020$ & $0.190\pm0.025$ & $0.385\pm0.020$ & $10.20\pm0.30$              & $15.50\pm1.00$\\
              & 30--40\% & $1.055\pm0.025$ & $0.805\pm0.101$ & $0.167\pm0.015$ & $0.365\pm0.010$ & $9.90\pm0.30$               & $17.00\pm0.50$\\
              & 40--50\% & $1.017\pm0.020$ & $0.779\pm0.016$ & $0.162\pm0.015$ & $0.335\pm0.005$ & $9.60\pm0.30$               & $18.50\pm0.50$\\
              & 50--60\% & $0.983\pm0.010$ & $0.754\pm0.018$ & $0.156\pm0.003$ & $0.335\pm0.005$ & $9.30\pm0.30$               & $20.00\pm0.50$\\
              & 60--70\% & $0.908\pm0.012$ & $0.699\pm0.014$ & $0.150\pm0.003$ & $0.285\pm0.005$ & $9.00\pm0.30$               & $23.00\pm0.30$\\
              & 70--80\% & $0.881\pm0.012$ & $0.679\pm0.013$ & $0.145\pm0.006$ & $0.265\pm0.005$ & $8.80\pm0.30$               & $24.00\pm0.30$\\
\hline
200 GeV $d$+Au & 0--20\%  & $0.972\pm0.015$ & $0.741\pm0.013$ & $0.163\pm0.006$ & $0.287\pm0.006$ & $9.60\pm0.50$ & $22.50\pm2.00$\\
               & 20--40\% & $0.965\pm0.010$ & $0.736\pm0.009$ & $0.153\pm0.006$ & $0.270\pm0.006$ & $9.30\pm0.50$ & $23.00\pm1.50$\\
               & 40--100\%& $0.924\pm0.010$ & $0.705\pm0.009$ & $0.148\pm0.007$ & $0.250\pm0.007$ & $9.00\pm0.50$ & $24.50\pm1.50$\\
\hline
2.76 TeV Pb+Pb & 0--5\%   & $0.889\pm0.023/1.957\pm0.142$ & $0.659\pm0.031/1.439\pm0.065$ & $0.310\pm0.060/0.170\pm0.010$ & $0.500\pm0.010/0.371\pm0.005$ & $22.50\pm1.50/11.00\pm0.10$ & $18.00\pm1.00/26.00\pm0.10$\\
               & 5--10\%  & $0.884\pm0.044/1.956\pm0.111$ & $0.656\pm0.020/1.439\pm0.043$ & $0.280\pm0.050/0.170\pm0.005$ & $0.500\pm0.010/0.370\pm0.005$ & $22.00\pm2.00/10.60\pm0.20$ & $20.00\pm1.00/26.30\pm0.20$\\
               & 10--20\% & $0.879\pm0.051/1.956\pm0.230$ & $0.653\pm0.025/1.439\pm0.071$ & $0.250\pm0.020/0.165\pm0.010$ & $0.480\pm0.010/0.370\pm0.005$ & $18.50\pm1.00/10.70\pm0.30$ & $20.50\pm1.50/26.50\pm0.50$\\
               & 20--30\% & $0.862\pm0.080/1.955\pm0.450$ & $0.642\pm0.023/1.439\pm0.038$ & $0.220\pm0.030/0.165\pm0.008$ & $0.450\pm0.010/0.365\pm0.005$ & $17.00\pm1.00/10.50\pm0.50$ & $21.50\pm1.00/26.80\pm0.80$\\
               & 30--40\% & $0.857\pm0.058/1.944\pm0.500$ & $0.638\pm0.015/1.428\pm0.027$ & $0.200\pm0.020/0.165\pm0.010$ & $0.450\pm0.010/0.360\pm0.007$ & $16.00\pm0.80/10.00\pm0.40$ & $23.50\pm1.50/27.00\pm0.50$\\
               & 40--50\% & $1.110\pm0.120              $ & $0.918\pm0.039              $ & $0.180\pm0.001              $ & $0.370\pm0.001              $ & $11.70\pm0.20$              & $26.50\pm0.50$\\
               & 50--60\% & $1.104\pm0.111              $ & $0.906\pm0.031              $ & $0.170\pm0.005              $ & $0.360\pm0.005              $ & $11.50\pm0.20$              & $27.00\pm0.20$\\
               & 60--70\% & $1.050\pm0.110              $ & $0.860\pm0.031              $ & $0.163\pm0.003              $ & $0.350\pm0.005              $ & $11.20\pm0.30$              & $27.50\pm0.50$\\
               & 70--80\% & $1.033\pm0.110              $ & $0.845\pm0.031              $ & $0.155\pm0.005              $ & $0.340\pm0.010              $ & $11.00\pm0.50$              & $28.00\pm1.00$\\
               & 80--90\% & $0.961\pm0.090              $ & $0.782\pm0.030              $ & $0.150\pm0.005              $ & $0.336\pm0.006              $ & $10.50\pm0.50$              & $29.00\pm1.00$\\
\hline
\end{tabular}}
\end{center}
\end{sidewaystable}
\begin{multicols}{2}

\begin{figure*}[!htb]
\begin{center}
\includegraphics[width=16cm]{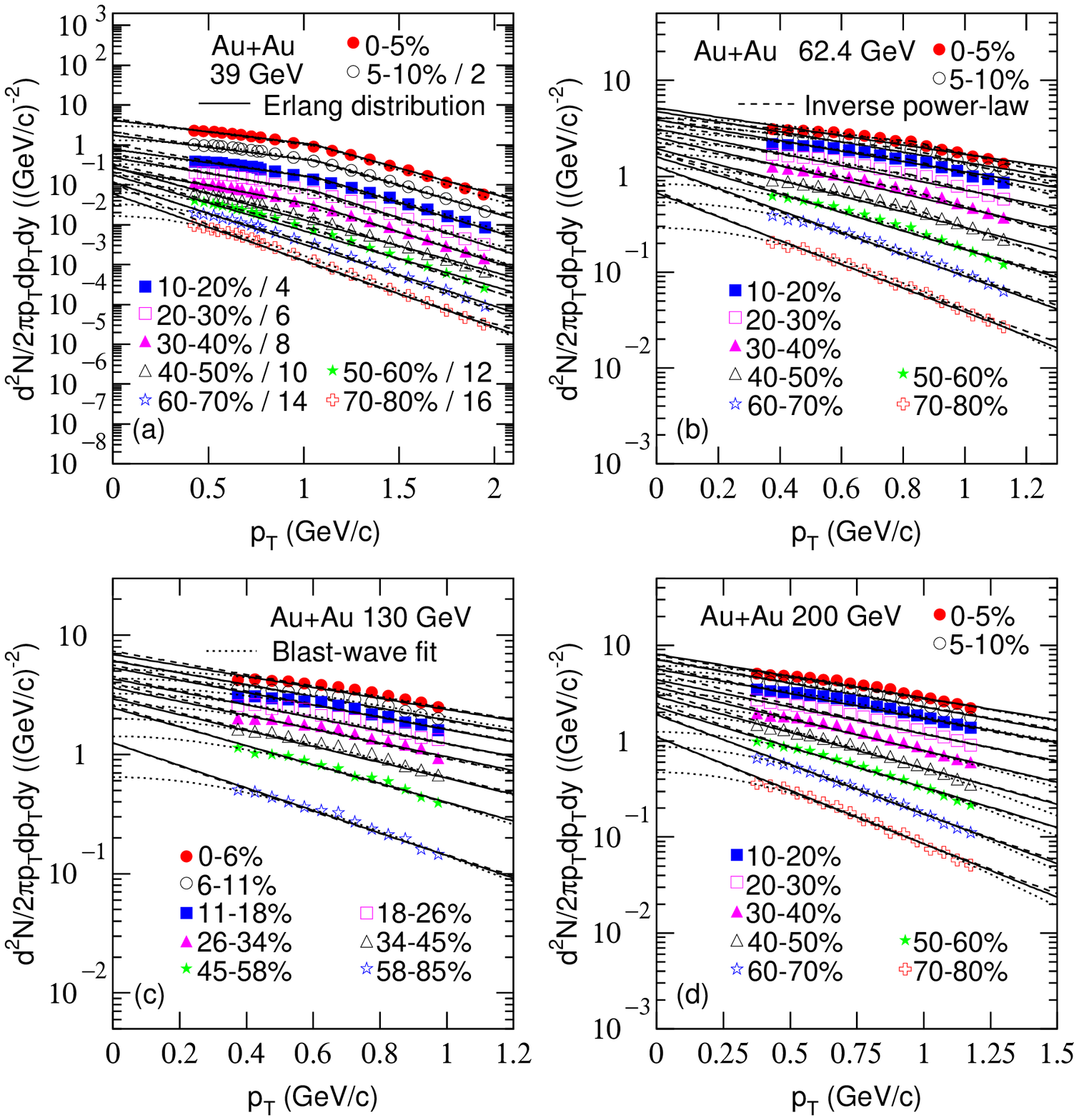}
\end{center}
{\small Fig. 4. Same as Fig. 3, but showing the results in
$|y|<0.5$ at (a) 39 GeV and in $|y|<1$ at (b) 62.4 GeV, (c) 130
GeV, and (d) 200 GeV, with respective centrality classes. The data
are taken from refs. [14, 41].}
\end{figure*}

\begin{figure*}[!htb]
\begin{center}
\includegraphics[width=16cm]{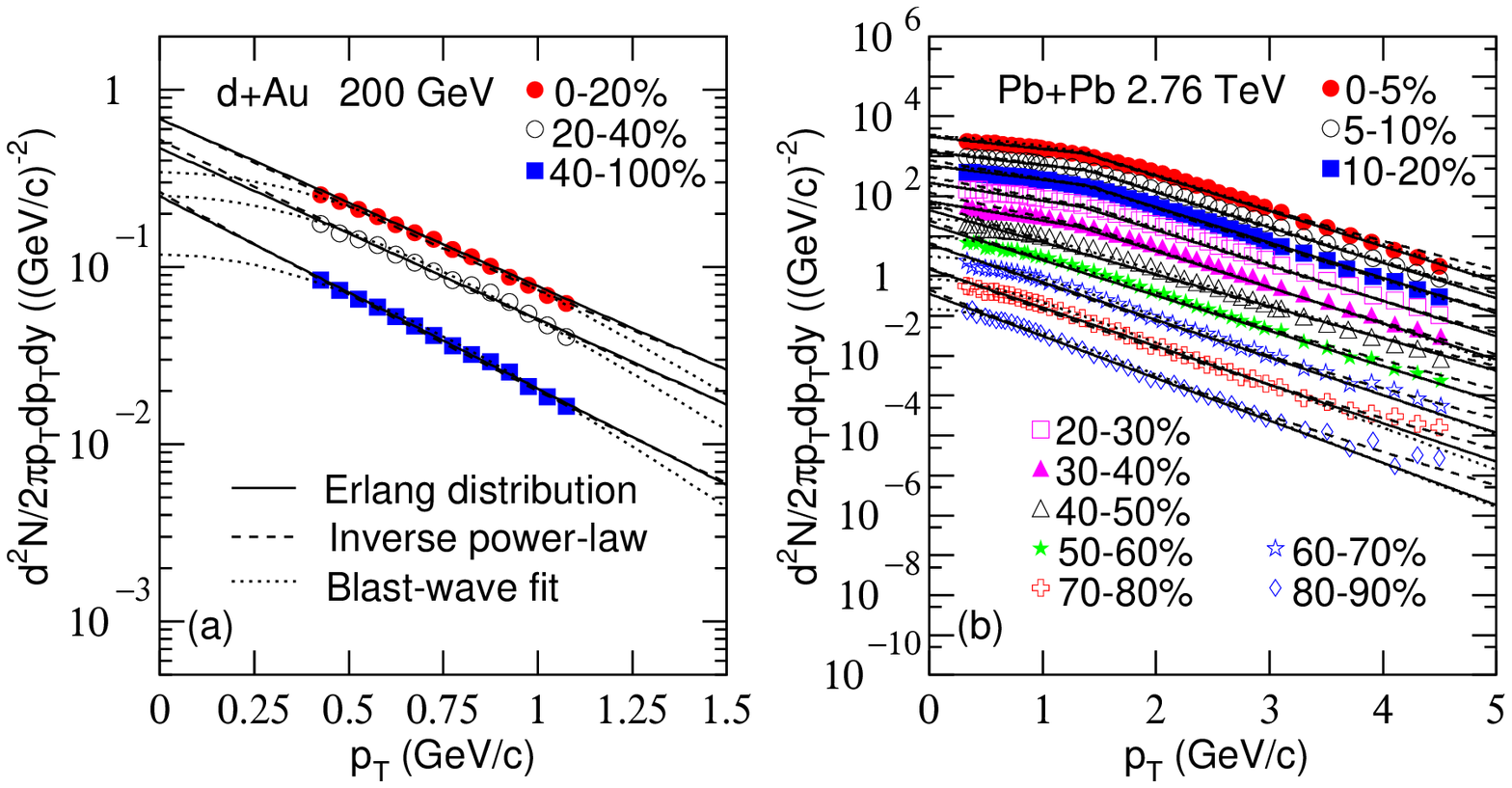}
\end{center}
{\small Fig. 5. Same as Fig. 3, but showing the results in $|y|<1$
in (a) $d$+Au collisions at 200 GeV and (b) Pb+Pb collisions at
2.76 TeV, with respective centrality classes. The data are taken
from refs. [14, 43].}
\end{figure*}

\begin{figure*}[!htb]
\begin{center}
\includegraphics[width=16cm]{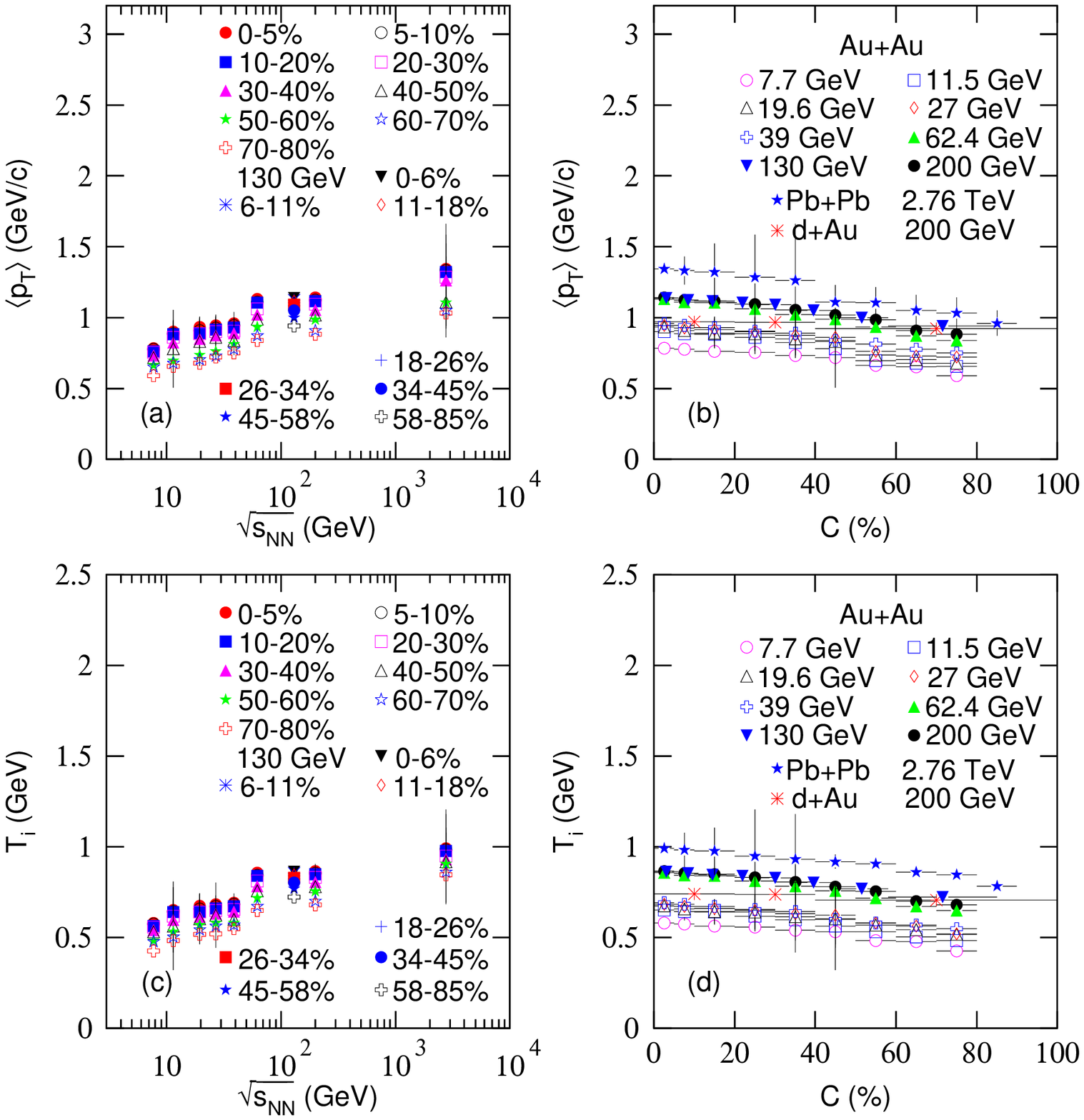}
\end{center}
{\small Fig. 6. Dependences of (a)(b) $\langle p_T\rangle$ and
(c)(d) $T_i$ on (a)(c) $\sqrt{s_{NN}}$ and (b)(d) $C$. The
different symbols represent the parameter values extracted from
Figs. 3--5 and listed in Table 2, where only the Erlang
distribution in the ranges of data available is used.}
\end{figure*}

\begin{figure*}[!htb]
\begin{center}
\includegraphics[width=16cm]{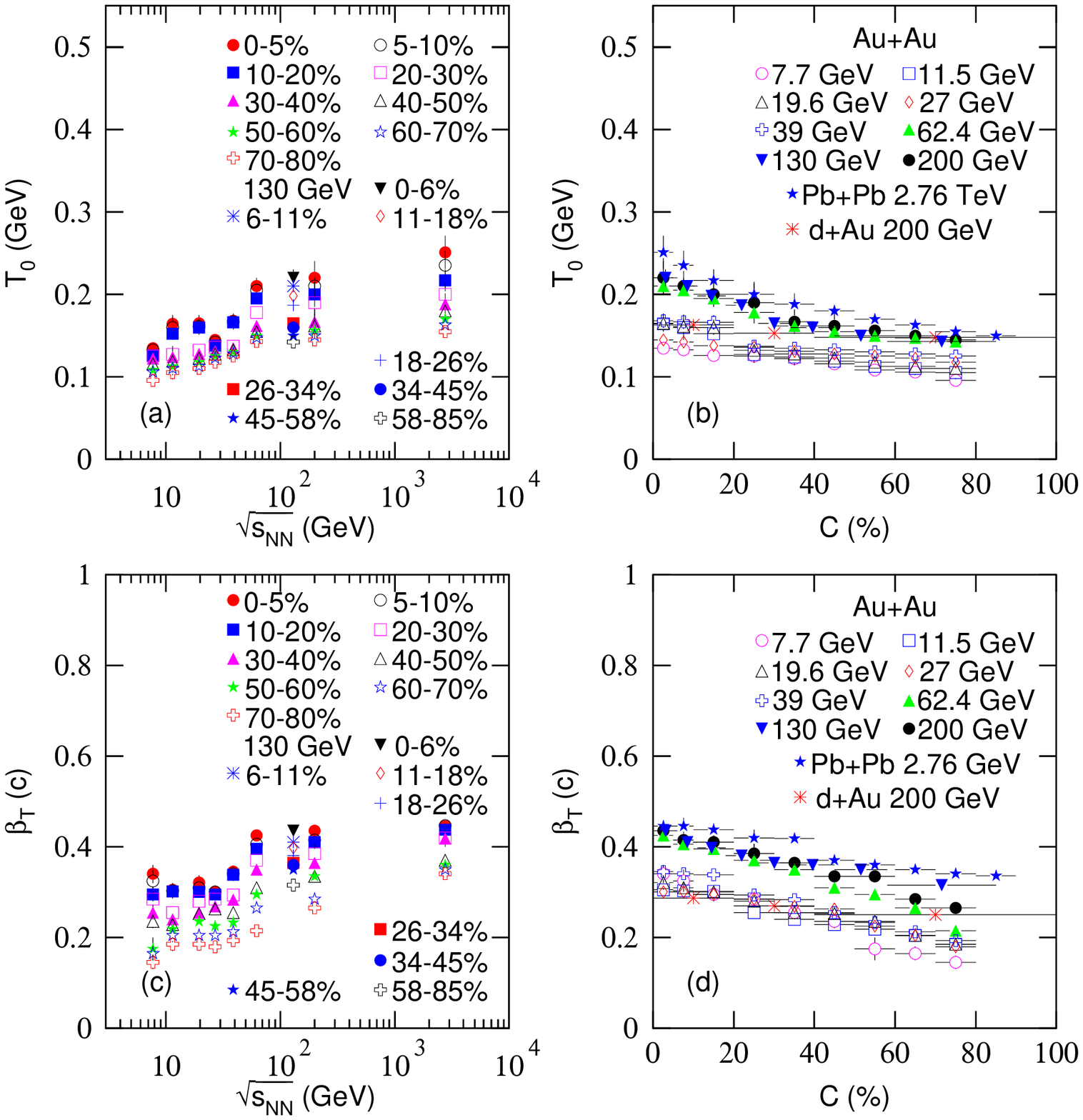}
\end{center}
{\small Fig. 7. Same as Fig. 6, but showing the dependences of
(a)(b) $T_0$ and (c)(d) $\beta_T$ on (a)(c) $\sqrt{s_{NN}}$ and
(b)(d) $C$. Only the blast-wave fit is used.}
\end{figure*}

\begin{figure*}[!htb]
\begin{center}
\includegraphics[width=16cm]{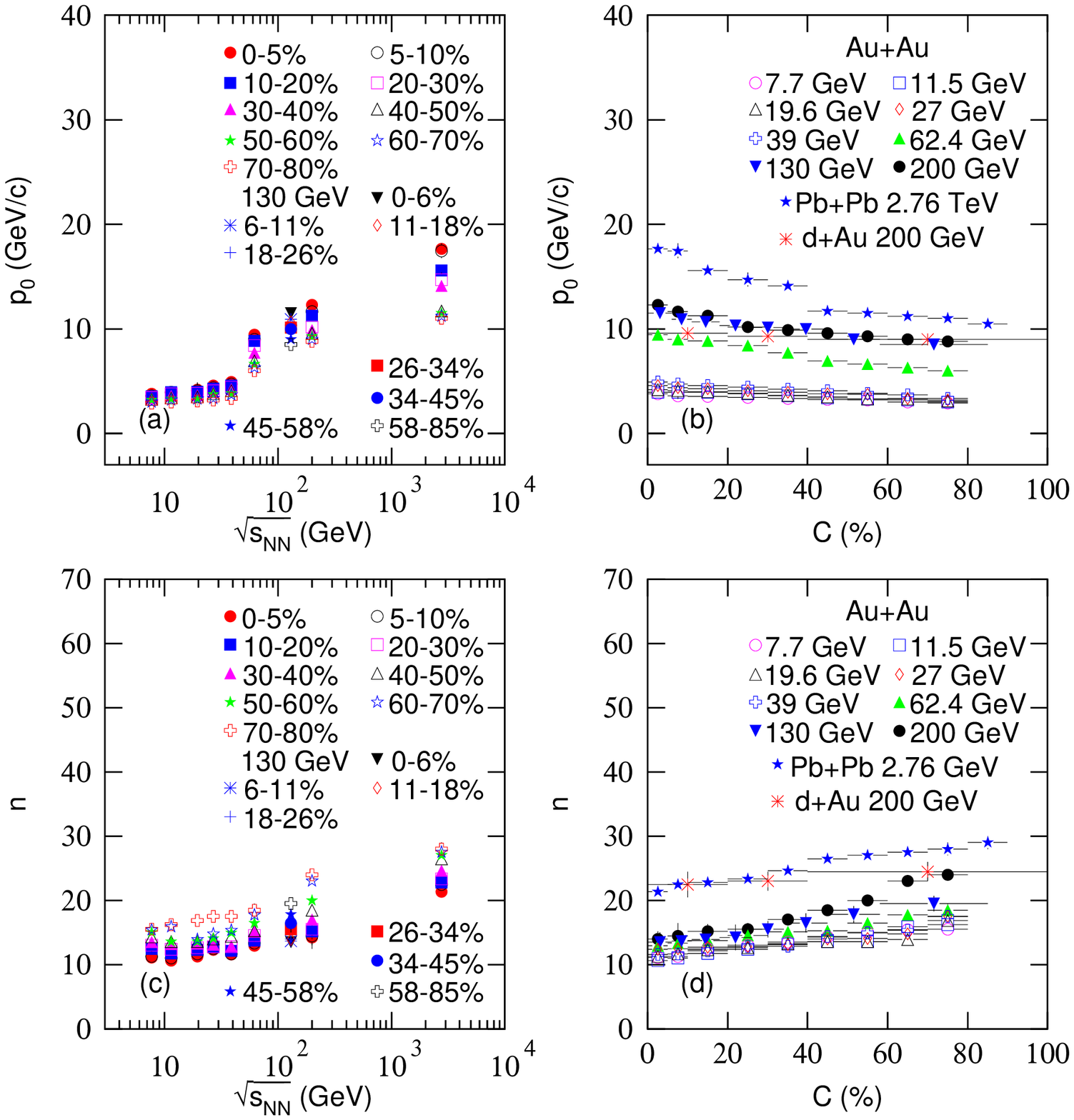}
\end{center}
{\small Fig. 8. Dependences of parameters (a,b) $T_0$, (c,d)
$\beta _T$ on energy in different centrality intervals (left
panel) and on centrality at different energies (right panel). The
different symbols represent the parameter values extracted from
fig. 3, 4, 5 and listed in table 2, where the Blast-wave model is
used for $\bar p$ spectra. The lines are our fitted results.}
\end{figure*}

{\section{Results and discussion}}

Figure 1 shows the momentum spectra, $d\sigma/dp$, of antiprotons
($\bar p$) produced in proton induced helium ($p$+He) collisions
at center-of-mass energy per nucleon pair $\sqrt{s_{NN}}=110$ GeV,
where $\sigma$ denotes the cross-section. Corresponding to the
left-upper, right-upper, and lower panels, the values of $p_T$ are
limited in 0.4--0.7, 0.7--1.2, and 1.2--2.8 GeV/$c$, respectively.
The circles presented in the three panels represent the
experimental data of $\bar p$ measured by the LHCb Collaboration
[39]. The data points are fitted by the Erlang distribution (the
solid curves) and the inverse power-law (the dashed curves)
respectively, which are obtained by using the Monte Carlo
calculation based on Eqs. (2) and (3) respectively. In the
calculations, the method of least square is used to obtain the
parameter values. The values of $\langle p_T\rangle$, $p_0$, and
$n$ are listed in Table 1 with $\chi^2$ per degree of freedom
(dof). One can see that the LHCb experimental data on antiproton
momenta in given transverse momentum ranges in $p$+He collisions
at $\sqrt{s_{NN}}=110$ GeV are approximately fitted by the Erlang
distribution and the inverse power-law, where the Monte Carlo
calculation is used to transform transverse momenta to momenta so
that the momentum distributions can be obtained.

Figure 2 shows the transverse momentum spectra, $d^2\sigma(\bar
pX)/dpdp_T$, of $\bar p$ produced in $p$+He collisions at 110 GeV,
where $X$ denotes other products except for $\bar p$. The symbols
represent the LHCb experimental data in different momentum ranges
[40], which are scaled by different amounts marked in the panels
(a)--(d). The data points are fitted by the Erlang distribution
Eq. (2) (the solid curves) and the inverse power-law Eq. (3) (the
dashed curves) respectively. In the fit, the Monte Carlo
calculation is used to select the momentum ranges and the method
of least square is used to obtain the parameter values. The values
of $\langle p_T\rangle$, $p_0$, and $n$ are listed in Table 1 with
$\chi^2$/dof. One can see that the LHCb experimental data on
antiproton transverse momenta in given momentum ranges in $p$+He
collisions at 110 GeV are approximately fitted by the Erlang
distribution and the inverse power-law, where the Monte Carlo
calculation is used to transform transverse momenta to momenta so
that the momentum ranges can be determined.

The transverse momentum spectra, $d^2N/2\pi p_Tdp_Tdy$, of $\bar
p$ produced in mid-rapidity interval ($|y|<0.5$) in gold-gold
(Au+Au) collisions at $\sqrt{s_{NN}}=$ (a) 7.7 GeV, (b) 11.5 GeV,
(c) 19.6 GeV, and (d) 27 GeV are presented in Fig. 3. Different
symbols represent the data measured by the STAR Collaboration in
the collision centrality classes of 0--5\%, 5--10\%, 10--20\%,
20--30\%, 30--40\%, 40--50\%, 50--60\%, 60--70\%, and 70--80\%
[41] and scaled by different amounts marked in the panels. The
solid, dashed, and dotted curves are our results fitted by using
the Erlang distribution Eq. (2), the inverse power-law Eq. (3),
and the blast-wave fit Eq. (5), respectively. The method of least
square is used to obtain the parameter values which are listed in
Table 2, where only $T_i$ obtained from Eq (2) in the range of
data available are listed. In some cases, a two-component
superposed by usual step function is used [42]. The parameter
values listed in Table 2 are then averaged by weighting the two
components. Figure 4 is the same as Fig. 3, but showing the
results in Au+Au collisions at $\sqrt{s_{NN}}=$ (a) 39 GeV with
$|y|<0.5$ and (b) 62.4 GeV, (c) 130 GeV, and (d) 200 GeV with
$|y|<1$, with respective centrality class shown in the panels. The
data are taken from refs. [14, 41]. Figure 5 is also the same as
Fig. 3, but showing the results in $|y|<1$ in (a) deuton-gold
($d$+Au) collisions at 200 GeV and (b) lead-lead (Pb+Pb)
collisions at 2.76 TeV with respective centrality class. The data
are taken from refs. [14, 43]. One can see that the STAR and LHCb
experimental data on antiproton transverse momenta in different
centrality classes in Au+Au, $d$+Au, and Pb+Pb collisions at high
energies are approximately fitted by the Erlang distribution, the
inverse power-law, and the blast-wave fit, though in some cases
the two-component is needed.

Figure 6 shows the dependences of parameters (a)(b) $\langle
p_T\rangle$ and (c)(d) $T_i$ on (a)(c) collision energy
($\sqrt{s_{NN}}$) in different centrality classes and (b)(d) event
centrality ($C$) at different energies. The different symbols
represent the parameter values extracted from Figs. 3--5 and
listed in Table 2, where only the Erlang distribution in the
ranges of data available is used. In the ranges of data available,
other two fits present similar results to the Erlang distribution.
One can see that $\langle p_T\rangle$ and $T_i$ increase slightly
with the increases of collision energy and event centrality, where
the centrality 0-5\% is the largest in the data samples cited in
this work.

Figure 7 is the same as Fig. 6, but showing the dependences of
parameters (a)(b) $T_0$ and (c)(d) $\beta_T$ on (a)(c)
$\sqrt{s_{NN}}$ and (b)(d) $C$. The different symbols represent
the parameter values extracted from Figs. 3--5 and listed in Table
2, where only the blast-wave fit is used. One can see that $T_0$
and $\beta_T$ increase slightly with the increases of collision
energy and event centrality. The trends of $T_0$, $\beta_T$,
$\langle p_T\rangle$, and $T_i$ are consistent with each other.
These results are natural due to the fact that the system at high
energy and with central centrality stays at the state with high
degrees of excitation and expansion, in which large energy is
deposited.

Figure 8 is the same as Fig. 6, but showing the dependences of
parameters (a)(b) $p_0$ and (c)(d) $n$ on (a)(c) $\sqrt{s_{NN}}$
and (b)(d) $C$. The different symbols represent the parameter
values extracted from Figs. 3--5 and listed in Table 2, where only
the inverse power-law is used. One can see that, similar to other
parameters ($T_0$, $\beta_T$, $\langle p_T\rangle$, and $T_i$)
discussed above, $p_0$ also increases (slightly) with the
increases of collision energy and event centrality. That is, $p_0$
also describes the excitation and expansion degrees of emission
source, which results in large $p_0$ at high energy and in central
collisions. Although $n$ increases slightly with the increase of
collision energy, it decreases slightly with the increase of event
centrality. We think that $n$ also describes the contribution
fraction of hard scattering process, which results in large $n$ at
high energy and in peripheral collisions.

There are fluctuations in the excitation functions (energy
dependences) of considered parameters. These fluctuations can be
regarded as the statistical fluctuations. To smooth these
statistical fluctuations, more analyses are needed in future. In
particular, more analyses are needed at energies below a few GeV
which is even below the energy range of beam energy scan program
[41]. We are very interested in this energy range due to the fact
that it possibly contains the critical energy of phase
transformation from hadronic matter to quark-gluon plasma. The
excitation functions of some parameters are expected to appear
with the minimum, maximum, corner, saturation, and/or limitation.
The starting points of saturation and limitation are particularly
worth to take attention.

Before conclusions, we would like to point out the mass dependence
of main parameters [34, 42, 44, 45]. Generally, $\langle
p_T\rangle$, $T_i$, and $T_0$ increase with the increase of
particle mass due to heavier particle corresponding to larger
energy deposition. Contrarily, $\beta_T$ decreases with the
increase of particle mass due to heavier particle having larger
inertia. Although the absolute values of some parameters are
model-dependent, the relative sizes are considerable. The average
parameter can be obtained by weighting different yields of various
particles. The weighted average of parameter values for various
particles can be regarded as the mass-independent parameter value
for given collisions. If the mass-independent parameter means
simultaneous production and freeze-out, the mass-dependent
parameter implies non-single scenario [46, 47].

It is well known that nowadays the Tsallis distribution [48--50]
is quite of use and seems to be very successful. In our previous
work [4, 34, 42, 44, 45, 51, 52], we have used the Tsallis
distribution and related functions to analyze the particle
production in high energy collisions. To express the variousness
of fit functions, we have used other functions in this work. It is
shown that we may use different functions to extract some main
parameters such as $\langle p_T\rangle$, $T_i$, $T_0$, and
$\beta_T$. Among these parameters, $\langle p_T\rangle$ and $T_i$
are model-independent, while $T_0$ and $\beta_T$ are
model-dependent. We hope to structure a model-independent method
to extract $T_0$ and $\beta_T$ in the near future.
\\

{\section{Conclusions}}

To conclude, the momentum or transverse momentum spectra of $\bar
p$ produced at mid-rapidity in $p$+He, Au+Au, $d$+Au, and Pb+Pb
collisions over an energy range from a few GeV to a few TeV have
been analyzed by the Erlang distribution, the inverse power-law
(the Hagedorn function), and the blast-wave fit. In some cases,
the usual step function is used to superpose the two-component
distribution. The model results are in agreement with the
experimental data of the STAR, ALICE, and LHCb Collaborations. The
values of related parameters are extracted from the fit process
and the excitation functions of these parameters are obtained.

The excitation functions of parameters $\langle p_T\rangle$,
$T_i$, $T_0$, $\beta_T$, and $p_0$ increase (slightly) from a few
GeV to a few TeV and from peripheral to central collisions. These
trends render that these parameters describe the excitation and
expansion degrees of the system. At high energy and in central
collisions, large collision energy is deposited in the system,
which results in high excitation and expansion degrees. The
excitation function of parameter $n$ shows a slight increase from
a few GeV to a few TeV and a slight decrease from peripheral to
central collisions. These trends render that the parameter $n$
also describes the contribution fraction of hard scattering
process. At high energy and in peripheral collisions, the hard
scattering process happens with large probability.
\\
\\
{\bf Data availability}

The data used to support the findings of this study are included
within the article and are cited at relevant places within the
text as references.
\\
\\
{\bf Compliance with ethical standards}

The authors declare that they are in compliance with ethical
standards regarding the content of this paper.
\\
\\
{\bf Conflict of Interest}

The authors declare that they have no conflict of interest
regarding the publication of this paper.
\\
\\
{\bf Acknowledgments}

This work was supported by the National Natural Science Foundation
of China under Grant Nos. 11575103 and 11847311, the Scientific
and Technological Innovation Programs of Higher Education
Institutions in Shanxi (STIP) under Grant No. 201802017, the
Shanxi Provincial Natural Science Foundation under Grant No.
201701D121005, the Fund for Shanxi ``1331 Project" Key Subjects
Construction
\\

\end{multicols}
\end{document}